\documentclass{PoS}

\PoS{PoS(LAT2005)351}

\usepackage{graphicx}
\usepackage{amsmath,amssymb}

\title{$\Delta I = 3/2$ kaon weak matrix elements with non-zero total momentum }

\ShortTitle{$\Delta I = 3/2$ kaon weak matrix elements with non-zero total momentum }

\author{\speaker{Takeshi Yamazaki} for the RBC Collaboration
\\
        RIKEN BNL Research Center, Brookhaven National Laboratory\\
	Upton, NY 11973, USA\\
        E-mail: \email{tyamazaki@bnl.gov}}

\abstract{
We present preliminary results for the $\Delta I = 3/2$ kaon decay
matrix elements using domain wall fermions and the DBW2 gauge action
at one coarse lattice spacing corresponding to $a^{-1} = 1.3$ GeV. 
We employ an extention of the Lellouch and L\"uscher formula
for non-zero total momentum to extract the infinite volume,
center-of-mass system decay amplitudes.  
We compare the result of $\mathrm{Re}A_2$ with
previous results calculated by several methods, and experiment.
We also show the $I=2$ $\pi\pi$ scattering phase shift.
}

\FullConference{XXIIIrd International Symposium on Lattice Field Theory\\
		 25-30 July 2005\\
		 Trinity College, Dublin, Ireland}

\begin{document}

\section{Introduction}

Non-leptonic kaon decay includes interesting subjects
such as, the $\Delta I = 1/2$ selection rule and the CP violation parameter
$\varepsilon^\prime / \varepsilon$.
However, it is difficult to calculate 
the $K \to \pi\pi$ weak matrix element directly on lattice
due to difficulty of calculation of the two-pion state in finite volume,
which was pointed out by Maiani and Testa~\cite{MT}.
So far many groups employ an indirect method~\cite{indirect_method} 
using Chiral perturbation theory (ChPT)
to avoid the difficulty.
In the indirect method $K\to\pi\pi$ process is reduced to $K\to\pi$ and 
$K\to 0$ processes.
Since in the decay process the final state interaction of the two-pion
is expected to play an important role, 
the reduction may cause systematic errors.

For the direct method, where the two-pion state is calculated on lattice,
we need complicated calculations and analyzes, 
{\it e.g.}, diagonalization of a matrix of correlation functions~\cite{LW},
to treat the two-pion state with non-zero relative momentum on lattice.
However, recently Kim~\cite{CK} reported 
an exploratory study of the direct calculation 
with H-parity boundary conditions,
where complicated analyzes are not required.
He succeeded to extract the two-pion state with non-zero relative momentum
from the ground state contribution of correlation functions,
because the two-pion state with zero momentum is prohibited by 
the boundary condition.
In the non-zero total momentum (Lab) system, $|\vec{P}|\ne 0$,
the ground state of the two-pions is $|\pi(0)\pi(\vec{P})\rangle$,
which is related to the two-pion state with the non-zero relative momentum
in center-of-mass (CM) system.
Thus, we can extract the two-pion state with non-zero momentum from
the ground state contributions as well as
in the H-parity boundary case.
However, we cannot naively apply the formula proposed by
Lellouch and L\"uscher (LL)~\cite{LL} to the calculation of the Lab system,
because the original formula is derived in the CM system.
Before the simulation,
we have to modify the LL formula to connect the finite volume decay amplitude
in the Lab system to that of the CM system in infinite volume.
Very recently, Kim {\it et al.}~\cite{KSS} and 
Christ {\it et al.}~\cite{CKY} suggested the formula
which is an extension of the LL formula for the Lab system calculation.
Here we attempt to apply this extended formula to the calculation of 
the $\Delta I = 3/2$ kaon weak matrix elements with domain wall fermions
and the DBW2 gauge action at a single coarse lattice spacing.

\section{Methods}

Two groups~\cite{KSS,CKY} proposed a formula between a decay amplitude 
in infinite $|A|$ (CM system) and finite volume $|M|$ (Lab system),
\begin{equation}
| A |^2 = 8\pi \gamma^2 \left(\frac{E_{\pi\pi}}{p}\right)^3
\left\{ p^\prime \frac{ \partial \delta }{ \partial p^\prime }
      + p^\prime \frac{ \partial \phi_{\vec{P}} }{ \partial p^\prime }
\right\}_{p^\prime=p} | M |^2,
\label{eq_Lab_Formula}
\end{equation}
where $\gamma$ is a boost factor, 
$E_{\pi\pi}$ is the two-pion energy in the CM system,
$p^2 = E_{\pi\pi}^2/4 - m_\pi^2$,
and $\delta$ is the scattering phase shift
of the final state interaction.
The function $\phi_{\vec{P}}$ with $\vec{P}$ being the total momentum, 
derived by Rummukainen and Gottlieb~\cite{RG},
is defined by
\begin{equation}
\tan \phi_{\vec{P}}(q) = 
- \frac{\gamma q\pi^{3/2}}{Z^{\vec{P}}_{00}(1;q^2;\gamma)},
\end{equation}
where $q^2 = ( p L / 2\pi )^2$, and
\begin{equation}
Z_{00}^{\vec{P}}(1;q^2;\gamma) = 
\frac{1}{\sqrt{4\pi}}
\sum_{\vec{n}\in \mathbb{Z}^3}\frac{1}{n_1^2+n_2^2+\gamma^{-2}(n_3+1/2)^2-q^2},
\label{eq_phi}
\end{equation}
in the $\vec{P} = (0, 0, 2\pi/L)$ case.
The formula eq.~(\ref{eq_Lab_Formula})
is valid only for on-shell decay amplitude,
{\it i.e.} $E_{\pi\pi} = m_K$ as is LL formula~\cite{LL}.

We calculate the four-point function for $\Delta I = 3/2$ $K\to\pi\pi$ decay
with total momentum $\vec{P} = \vec{0}$ and $(0, 0, 2\pi/L)$.
The four-point function $G_i(t)$ is defined by
\begin{equation}
G_i(t) = 
\langle 0 | 
\pi^+\pi^-(\vec{P},t_\pi) O^{3/2}_i(t) K^0(\vec{P},t_K) 
| 0 \rangle,
\end{equation}
where the operators $O^{3/2}_i$ are lattice operators entering
$\Delta I = 3/2$ weak decays
\begin{eqnarray}
O^{3/2}_{27,88} &=& 
(\overline{s}^ad^a)_L
\left[(\overline{u}^bu^b)_{L,R}-(\overline{d}^bd^b)_{L,R}\right]
       + (\overline{s}^au^a)_L(\overline{u}^bd^b)_{L,R}\\
O^{3/2}_{m88} &=& 
(\overline{s}^ad^b)_L
\left[(\overline{u}^bu^a)_{R}-(\overline{d}^bd^a)_{R}\right]
       + (\overline{s}^au^b)_L(\overline{u}^bd^a)_{R}
\end{eqnarray}
with $(\overline{q}q)_{L,R} = \overline{q}\gamma_\mu(1\pm\gamma_5)q$
and $a,b$ being color indices.
$O^{3/2}_{27}$ and $O^{3/2}_{88}$
are the operators in the (27,1) and (8,8) representations of 
$SU(3)_L\otimes SU(3)_R$ with $I=3/2$, respectively.
$O^{3/2}_{m88}$ equals $O^{3/2}_{88}$ with its color summation
changed to cross the two currents.
We also calculate four-point function for two-pions and the 
two-point function of a kaon with zero and non-zero total momenta,
to obtain those energies and amplitudes.

We employ the domain wall fermion action with the domain wall
height $M=1.8$, the fifth dimension length $L_s = 12$
and the DBW2 gauge action with $\beta = 0.87$ corresponding
to $a^{-1} = 1.3$ GeV.
The lattice size is $L^3 \cdot T = 16^3 \cdot 32$, 
where the spatial volume corresponds to about 2.4 fm.
We use four light quark masses, $m_u = 0.015, 0.03, 0.04$ and 0.05,
for the chiral extrapolation of the decay amplitudes,
and six strange quark masses, $m_s = 0.12, 0.18, 0.24, 0.28, 0.35$ and 0.44,
for the interpolation of the amplitudes to the on-shell point.
We fix the two-pion operator at $t_\pi = 0$,
while we employ three source points $t_K = 16, 20$ and $25$ 
for the kaon operator to investigate the $t_K$ dependence of 
the statistical error of the Lab system decay amplitude 
and to check the consistency of these results.
The numbers of configuration used are 
111 for $t_K = 16$ and $20$, and 100 for $t_K = 25$.
A quark propagator is calculated by averaging 
quark propagators with periodic and anti-periodic boundary conditions for 
the time direction to obtain a propagator with $2T$ periodicity.
We also use a wall source with Coulomb gauge fixing for the quark propagators.

\section{Results}

\begin{figure}[!t]
\begin{center}
\scalebox{0.30}[0.27]{
\rotatebox{270}{
\includegraphics{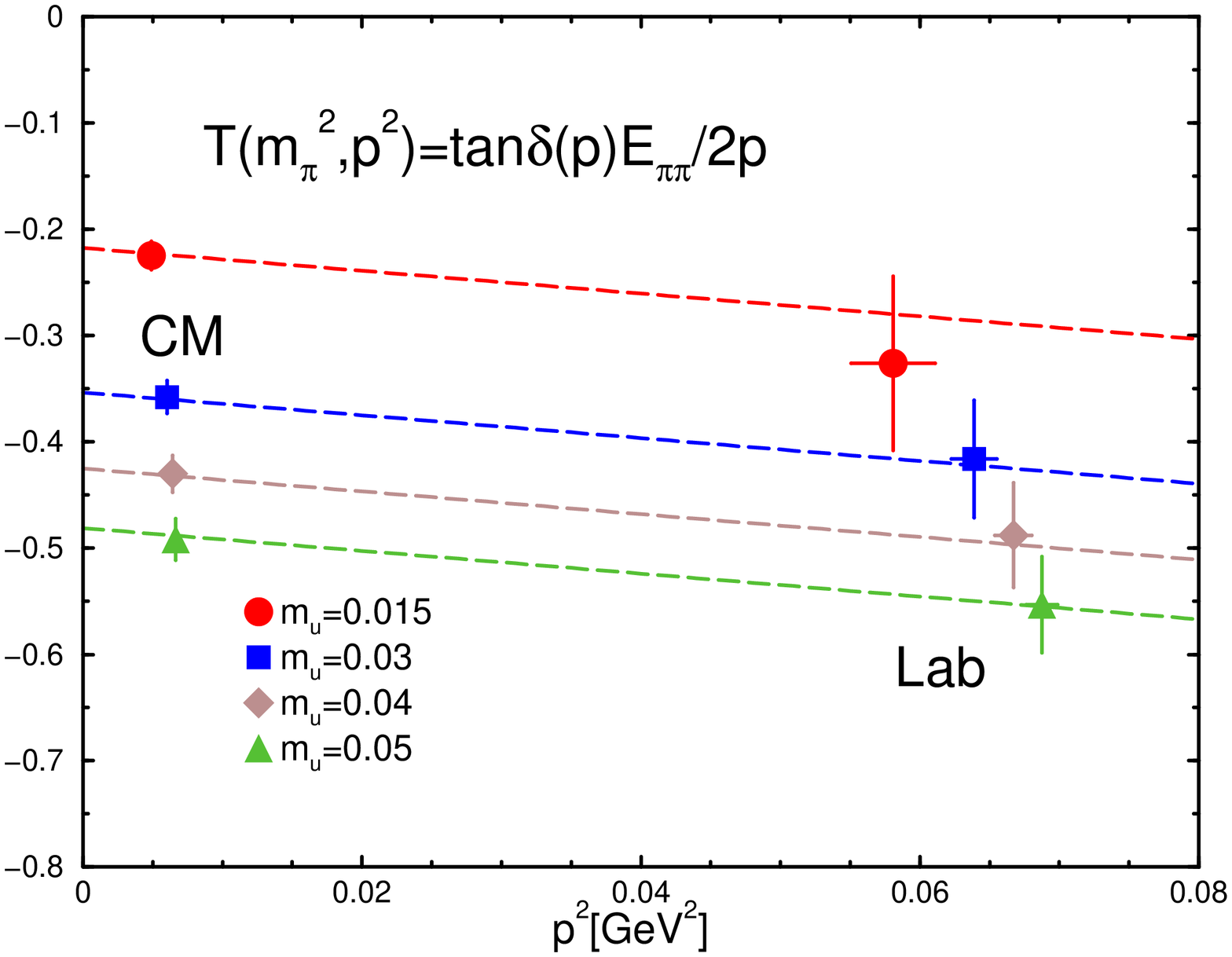}}}
\scalebox{0.30}[0.27]{
\rotatebox{270}{
\includegraphics{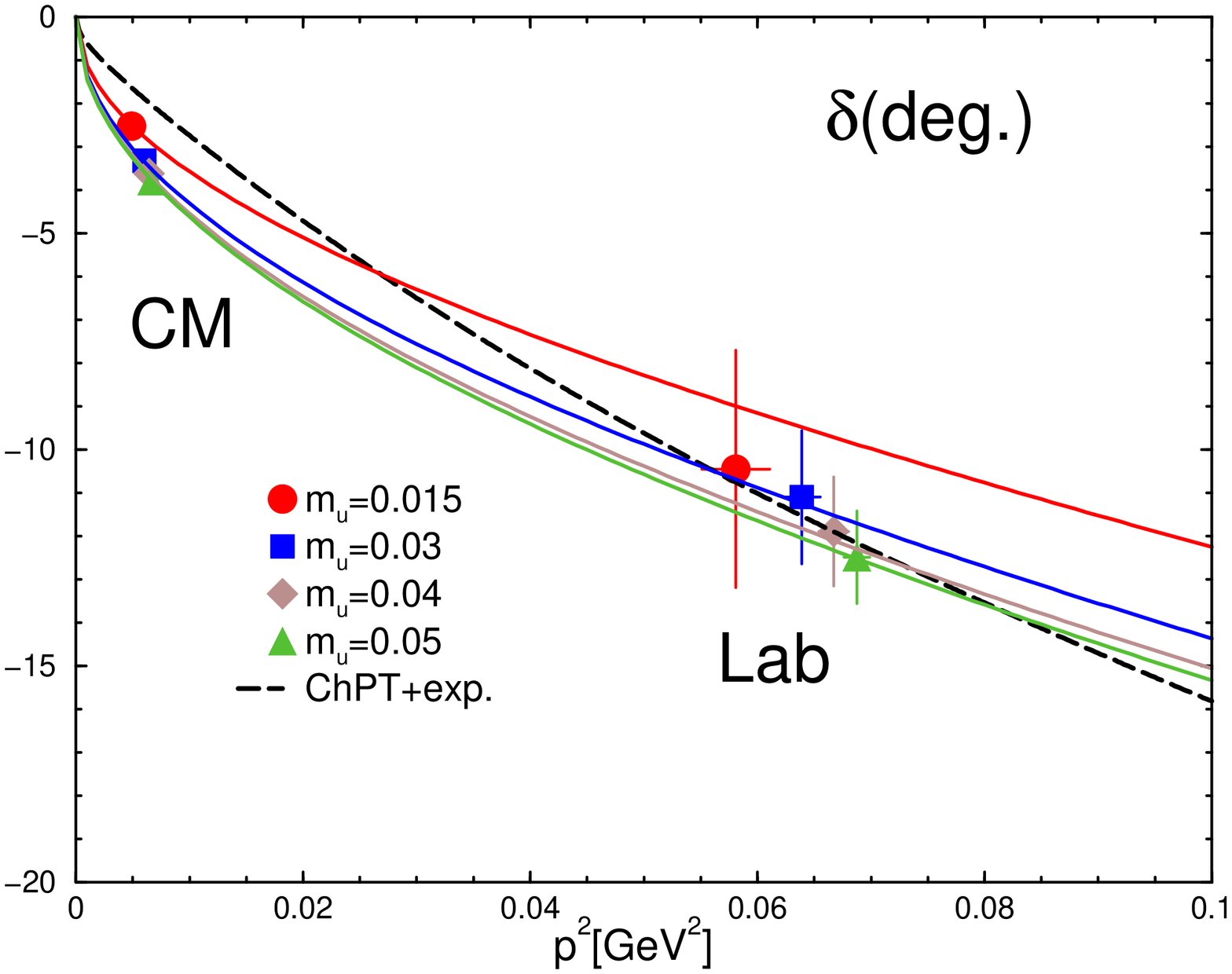}}}
\vspace{-5mm}
\end{center}
\caption{Measured values of $T$, defined in eq.~(\protect\ref{eq_T}),
and $\delta$ obtained from CM and Lab calculations.
Solid lines are fitting results.
\label{fig_delta}}
\end{figure}

In order to utilize the eq.~(\ref{eq_Lab_Formula}) 
we need to measure derivatives
of the scattering phase shift for the final state and 
evaluate the function $\phi_{\vec{P}}(q)$ defined in eq.~(\ref{eq_phi}).
The phase shift $\delta(p)$ is evaluated
from the two-pion energy by the finite volume formula~\cite{RG,FVF}.
We carry out a global fitting of
\begin{equation}
T(m_\pi,p) = \frac{ \tan \delta(p) }{ p }\cdot\frac{ E_{\pi\pi} }{ 2 }
\label{eq_T}
\end{equation}
with a polynomial function $a_{10} m_\pi^2 + a_{20} m_\pi^4 + a_{01} p^2$
to obtain the derivative of $\delta$ with respect to $p$.
The measured value of $T$ (left graph) and $\delta$ (right graph),
and their fitting results are plotted in Fig.~\ref{fig_delta}.
The prediction of $\delta$ from ChPT with experiment is also
shown in the right graph.

To determine off-shell decay amplitudes in finite volume,
we define ratio of correlation functions $R_i(t)$ for $i=27,88$ and $m88$ as
\begin{equation}
R_i(t) = 
\frac{ \sqrt{3} G_i(t) Z_{\pi\pi} Z_K }{ G_{\pi\pi}(t) G_K(t) },
\label{eq_ratio}
\end{equation}
where $G_{\pi\pi}(t)$ and $G_K(t)$ are $I=2$ two-pion four-point
function and kaon two-point function, respectively, and
$Z_{\pi\pi}$ and $Z_K$ are the overlap of the relevant operator with each state.
The factor $\sqrt{3}$ comes from changing basis to the isospin basis.
When these correlation functions are dominated by each ground state,
the ratio $R_i(t)$ will be a constant for those values of $t$.  We then determine
the off-shell amplitude in this region.

\begin{figure}[!t]
\begin{center}
\scalebox{0.27}[0.25]{
\rotatebox{270}{
\includegraphics{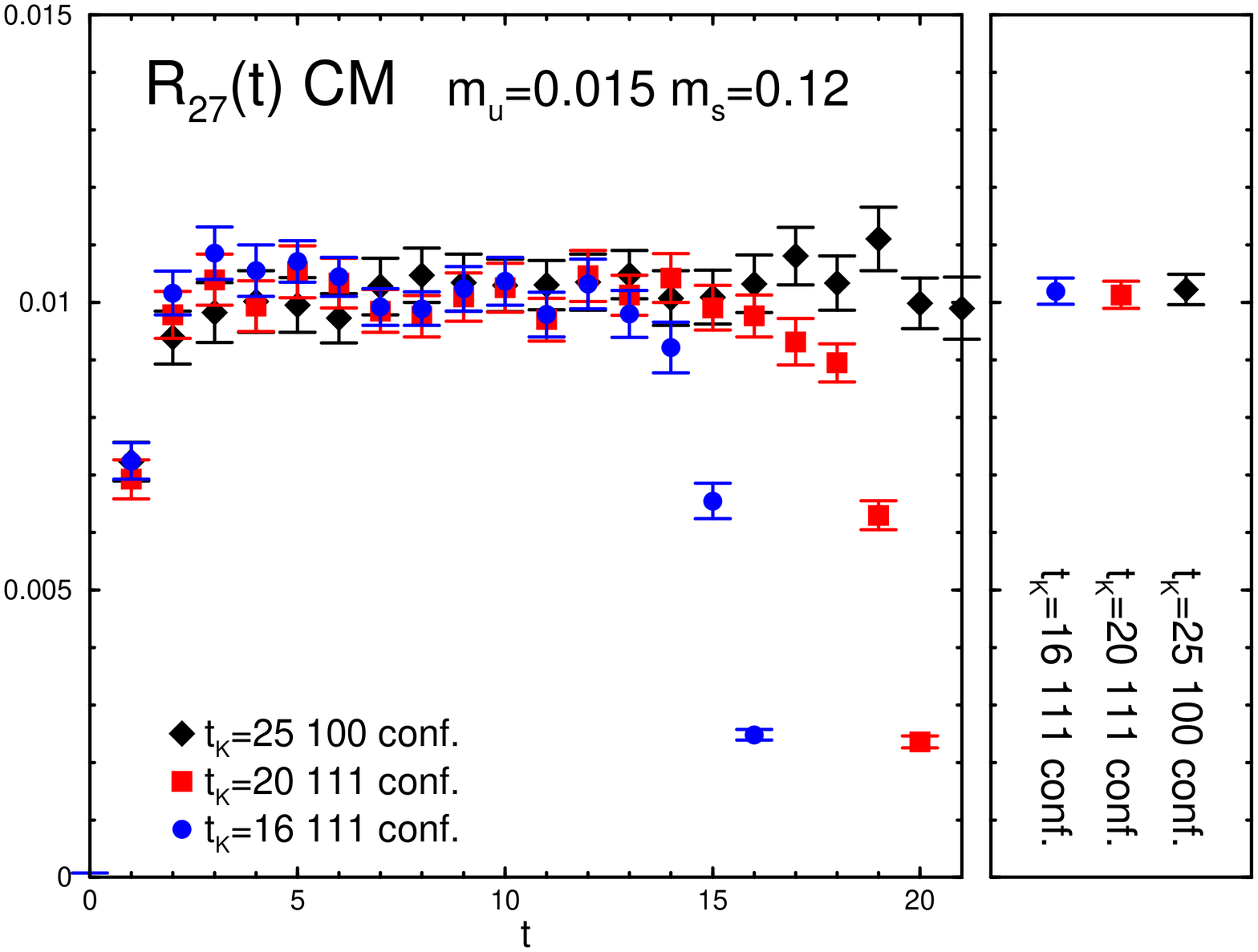}}}
\scalebox{0.27}[0.25]{
\rotatebox{270}{
\includegraphics{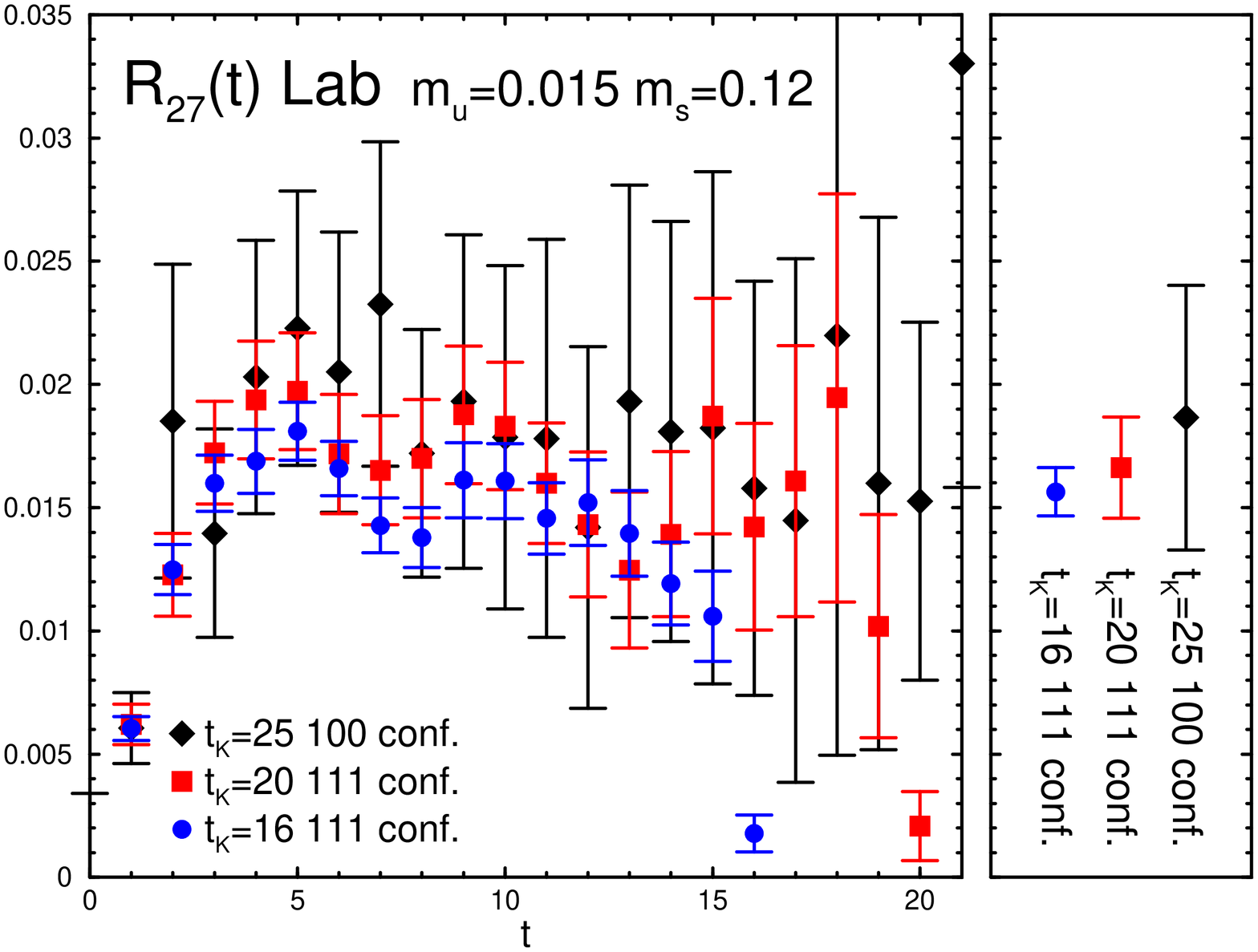}}}
\vspace{-5mm}
\end{center}
\caption{
$R_{27}(t)$, defined in eq.~(\protect\ref{eq_ratio}),
with different $t_K$ obtained from CM and Lab system calculations.
Small figures show averaged values in flat regions.
\label{fig_R}}
\end{figure}

The figure~\ref{fig_R} shows $R_{27}(t)$ for the lightest
pion mass and fixed strange quark mass obtained from the CM and Lab 
system calculations.
We plot the ratios with three different kaon source points $t_K$ in
the figure.
In the CM system, the ratios with different $t_K$
are reasonably flat and consistent with each other
in a region $t_{\pi} \ll t \ll t_K$.
The off-shell amplitudes are determined from averaged values in the 
flat region, whose values are presented in small figure.
The three averaged values are also consistent with each other.
In the Lab system case, the ratios are noisier than those in 
the CM case.
While the error with $t_K=25$ is very large,
the error decreases as $t_k$ decreases.
Thus, we choose the results with $t_K=16$ in the following analysis,
because the result is consistent with the others, 
and has the smallest error of the three cases.
We determine the on-shell amplitude in finite volume by 
a linear interpolation of the off-shell amplitude
with different strange quark masses at fixed light quark mass.
The decay amplitudes in infinite volume are obtained for 
the extended formula of eq.~(\ref{eq_Lab_Formula}) by
combining the on-shell amplitude and
the derivatives of $\delta$ and $\phi_{\vec{P}}$.

\begin{figure}[!h]
\begin{center}
\scalebox{0.29}[0.27]{
\rotatebox{270}{
\includegraphics{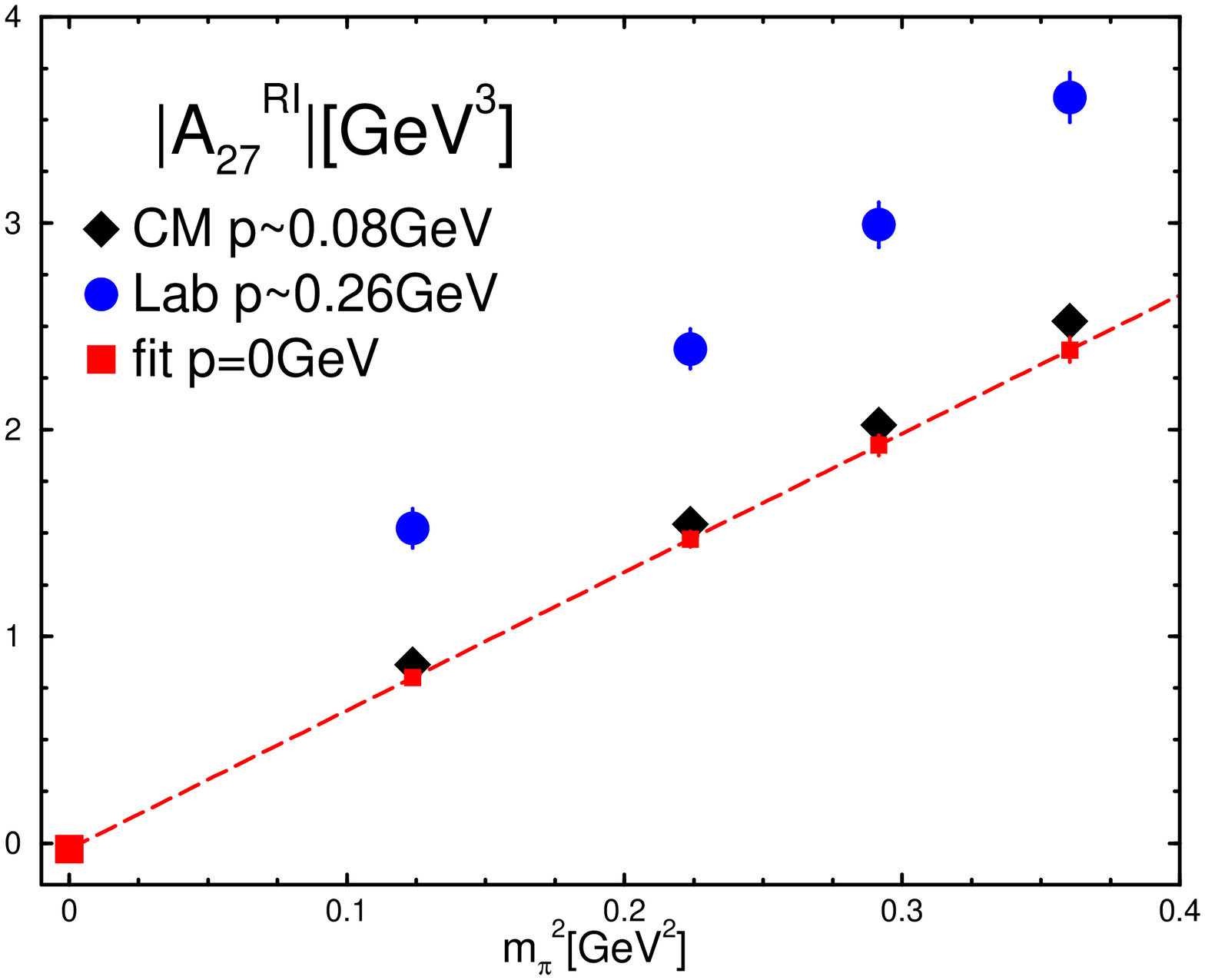}}}
\scalebox{0.24}[0.24]{
\rotatebox{270}{
\includegraphics{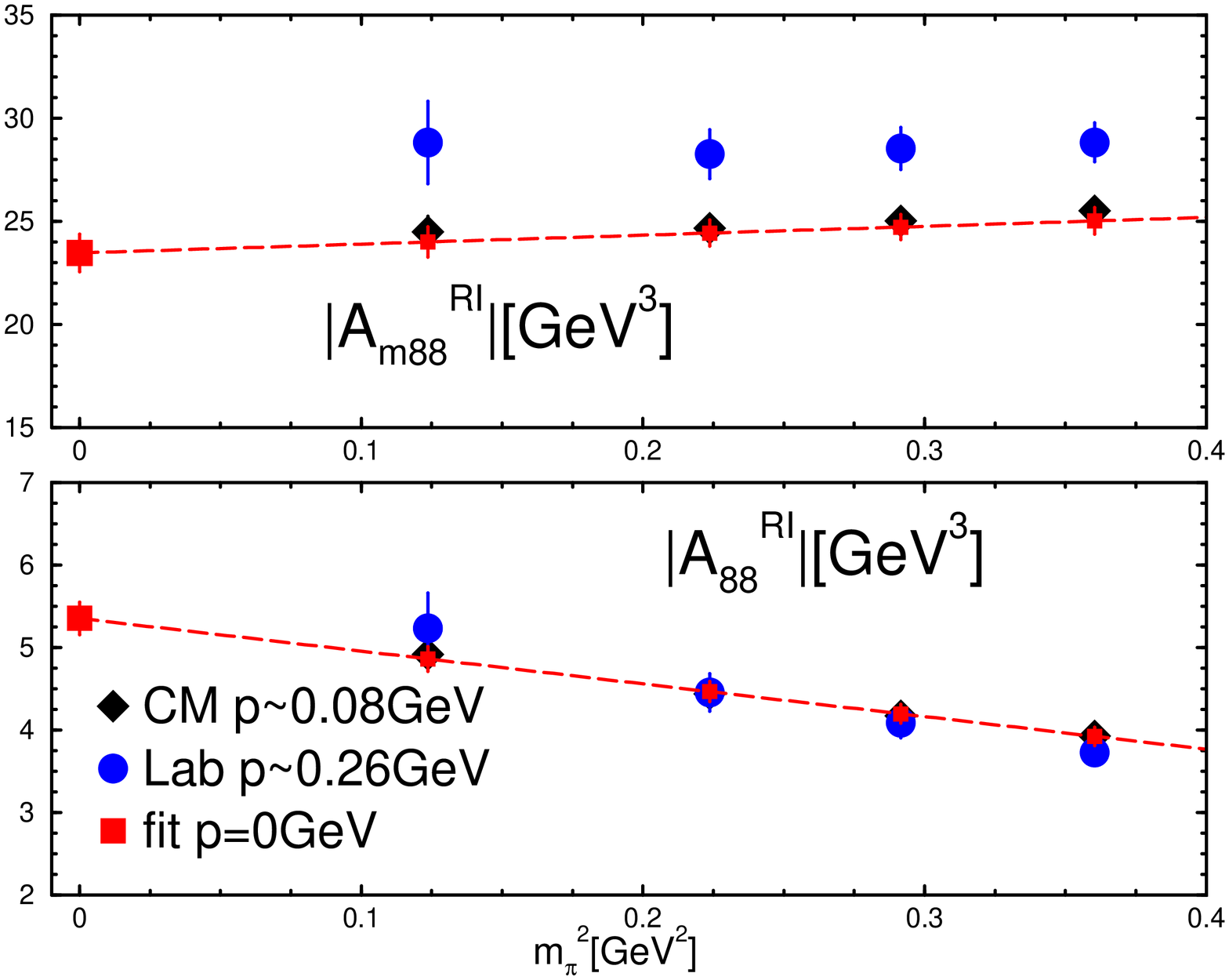}}}
\vspace{-5mm}
\end{center}
\caption{Weak matrix elements obtained from CM and Lab system calculations.
Squre symbols are fitting results with $p=0$.
\label{fig_wme}}
\end{figure}
\begin{figure}[!h]
\begin{center}
\scalebox{0.32}[0.32]{
\rotatebox{270}{
\includegraphics{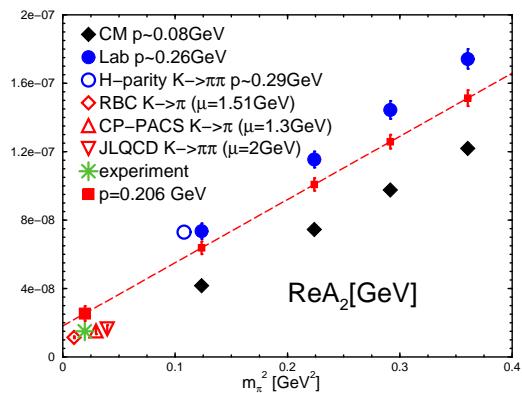}}}
\vspace{-5mm}
\end{center}
\caption{Measured $\mathrm{Re}A_2$ and result from previous works.
Square symbols denote fitting results.
\label{fig_ReA2}}
\end{figure}

The matching factors to determine the weak matrix elements
were previously calculated
in Ref.~\cite{CK} with the renormalization independent (RI) 
scheme by a non-perturbative method at the scale $\mu = 1.44$ GeV.
The weak matrix elements $|A_i^{\mathrm{RI}}|$ for $i=27,88$ and $m88$
in the RI scheme are plotted in Fig.~\ref{fig_wme}.
The weak matrix element of the 27 operator seems to vanish in the chiral limit
with $p^2 = 0$,
while the other elements remain a constant in the limit.
These trends in the pion mass dependence
are reasonably consistent with the prediction of ChPT~\cite{BGLLMPS}.
In order to investigate the $m_\pi$ and $p$ dependences,
we carry out a global fitting for each matrix element
for $m_\pi^2$ and $p^2$ assuming a simple polynomial form as
\begin{equation}
A_{00} + A_{10} m_\pi^2 + A_{11} m_\pi^2 p^2 + A_{01} p^2.
\label{fit_func}
\end{equation}
The fitting results at zero momentum for each pion mass
and at the chiral limit are represented by square symbols
in the figures.
The constant in the fitting result $A_{00}$ for the 27 operator 
is consistent with zero within the error, $A_{00}^{27} = -0.028(25)$, 
as expected.

We calculate $\mathrm{Re}A_2$ from the weak matrix elements 
with the Wilson coefficients evaluated by NDR scheme,
which is converted to RI scheme at the scale $\mu = 1.44$ GeV.
The $\mathrm{Re}A_2$ obtained from the CM and Lab system calculations
is shown in Fig.~\ref{fig_ReA2}.
We also plot the results previously obtained with 
H-parity boundary calculation~\cite{CK},
reduction method~\cite{CP-PACS_RBC}, and
direct calculation with ChPT~\cite{JLQCD}.
The results strongly depend on the pion mass, 
and also the momentum.
We estimate $\mathrm{Re}A_2$ at the physical point,
$m_\pi = 0.14$ GeV, $m_K = 0.498$ GeV and $p = 0.206$ GeV, 
by a global fit making the same polynomial assumption 
as for the case of the weak matrix elements in eq.~(\ref{fit_func}).
The square symbols in the figure denote 
the fitting results at $p = 0.206$ GeV.
The $\mathrm{Re}A_2$ at the physical point is $2.54(43)\times 10^{-8}$ GeV
which is 1.69(28) times larger than the experiment.
In order to understand the difference from the experiment,
we need a more detailed investigation of the systematic errors,
{\it e.g.},
finite volume effects, discretization errors, and quenching effect.

\section*{Acknowledgments}

We thank Changhoan Kim for his previous study upon which 
the present work is based, and also thank RIKEN BNL Research Center, 
BNL and the U.S. DOE for providing the facilities essential for 
the completion of this work.

\providecommand{\href}[2]{#2}\begingroup\raggedright

\end{document}